# Coherent multi-heterodyne spectroscopy using acousto-optic frequency combs

V. DURÁN,[1,2*] C. SCHNÉBELIN,[1] AND H. GUILLET DE CHATELLUS[1]

[1]*Univ. Grenoble Alpes, CNRS, LIPhy, F-38000 Grenoble, France*
[2]*GROC•UJI, Institut of New Imaging Technologies, Universitat Jaume I, 12070 Castelló, Spain*
*\*vduran@uji.es*

We propose and characterize experimentally a new source of optical frequency combs for performing multi-heterodyne spectrometry. This comb modality is based on a frequency shifting loop seeded with a CW monochromatic laser. The comb lines are generated by successive passes of the CW laser through an acousto-optic frequency shifter. We report the generation of frequency combs with more than 1500 mutually coherent lines, without resorting to non-linear broadening phenomena or external electronic modulation. The comb line spacing is easily reconfigurable from tens of MHz down to the kHz region. We first use a single acousto-optic frequency comb to conduct self-heterodyne interferometry with a high frequency resolution (500 kHz). By increasing the line spacing to 80 MHz, we demonstrate molecular spectroscopy on the sub-millisecond time scale. In order to reduce the detection bandwidth, we subsequently implement an acousto-optic dual-comb spectrometer with the aid of two mutually coherent frequency shifting loops. In each architecture, the potentiality of acousto-optic frequency combs for spectroscopy is validated by spectral measurements of hydrogen cyanide in the near-infrared region.

## 1. Introduction

Optical frequency combs (FC) have become key measurement tools in optical metrology and high-precision spectroscopy. They can be produced in a variety of ways, depending on the targeted application. Femtosecond mode-locked lasers are the most common sources of frequency combs, providing thousands of narrow teeth [1]. When femtosecond combs are fully stabilized they turn into extremely precise optical frequency rulers [2]. These capabilities have been exploited in diverse areas, including optical clocks, molecular spectroscopy and distance ranging [3-6]. However the line spacing, or free spectral range (FSR), is set by the length of the laser cavity and is usually limited to values below the GHz range. This can be a critical issue for important applications, such as astronomical spectrograph calibration [7] or optical telecommunications [8]. Another limitation of femtosecond FCs is the relative complexity of self-referencing techniques to control the comb absolute frequency. A different way of generating phase-locked FCs with controlled absolute frequency consists of seeding a high-finesse resonator with a CW laser: regularly spaced comb lines are generated by non-linear interactions in the resonator. Combs with FSR ranging from 1 GHz to 1 THz have been reported in microresonators [9, 10]. This technology also offers the possibility of on-chip integration [9], self-referencing [11] and promising prospects for microwave photonics [12], dual-comb spectroscopy [13] and coherent optical telecommunications [14, 15]. However, similarly to femtosecond frequency combs, the FSR is directly fixed by the length of the resonator.

Alternatively, a relatively simple manner of producing combs with flexible FSR (up to tens of GHz), and tunable central frequency, is based on the modulation of a continuous wave (CW) laser using electro-optic modulators [16]. The line spacing is set by the radio-frequency

(RF) function generator driving the modulators, while the central frequency is determined by the CW laser. Besides, the generation of so-called electro-optic FC is particularly simple with conventional telecommunications components. Apart from their versatility, electro-optic FCs have recently aroused great interest for conducting dual-comb spectroscopy, since the two combs that are involved in that technique can be produced from a single CW laser. This feature ensures a high degree of mutual coherence without resorting to sophisticated phase-locking schemes [17-19]. Beyond spectroscopy, electro-optic dual-comb interferometers have been applied to optical waveform characterization [20, 21], laser ranging [22] and vibrometry [23, 24]. However, despite the above advantages, electro-optic FCs suffer from a relatively low number of teeth (typically <100). Different techniques have been proposed to increase the number of spectral lines. Non-linear broadening in a highly nonlinear fiber has proven successful to increase the number of lines by one order of magnitude, enabling frequency-agile dual-comb spectroscopy [19] or optical waveform characterization over the entire C-band [25]. Indeed, two successive spectral broadening stages, combined with an *f*-2*f* self-referencing interferometer, have been applied to reach a bidirectional frequency conversion between the microwave and optical domains [26]. Alternatively, spectroscopy with hundreds of lines can be conducted when the electro-optic modulation is driven by step-recovery diodes [27] or phase bit sequences [28, 29]. This avoids increasing the complexity of the comb-based systems by the insertion of a nonlinear broadening stage. An electro-optic frequency comb driven by random phase modulation has been employed, for instance, to perform dynamic atomic spectroscopy with high-precision frequency resolution (10 MHz) [30]. The extension to molecular spectroscopy has been subsequently demonstrated with a hybrid dual-comb setup, including a mode-locked frequency comb [31]. However, in these phase modulation techniques the spectral bandwidth of the FC is ultimately limited by the bandwidth of the function generator (a few tens of GHz). Another approach that has been proposed to produce FCs from a CW laser combines gain-switching and optical injection locking in a semiconductor laser [32, 33]. This method makes it possible to achieve dual-comb generation (potentially from visible to mid-infrared), although the spectra thus obtained are still limited to tens of lines.

Here we propose and implement a simple source of optical FCs, providing more than 1000 mutually coherent lines from a single CW laser, with a FSR reconfigurable from the kHz to the tens of MHz range. Our technique makes no use of non-linear media, nor fast electronic function generator. It is based on a simple frequency-shifting loop (FSL) seeded by a CW laser [34-38]. The comb lines are produced by successive frequency shifts in the loop. This optical system can be implemented in a straightforward manner with an acousto-optic frequency shifter (AOFS). Notice that seeded FSLs have been previously employed for generating pulse trains with tunable and ultrahigh repetition rates [35], as well as for performing optical real-time Fourier transformations [37, 38]. These time-domain applications take advantage of the temporal Talbot effect, which is originated by the controllable spectral quadratic phase exhibited by the comb modes [36]. In the frequency domain, it has been shown that acousto-optic frequency shifting loops can generate relatively flat spectra composed of >1000 modes [39]. However, the use of this kind of comb sources for spectroscopic applications has remained so far unexplored. In this paper, we uncover the potentiality of acousto-optic FCs for multi-heterodyne spectroscopy with high spectral resolution. In a first set of experiments, we demonstrate self-heterodyne interferometry [28, 30] for measuring the transmission peaks of a Fabry-Pérot cavity with a sub-MHz resolution. Then, by extending the tooth spacing to tens of MHz, we report molecular spectroscopy on the sub-millisecond time scale, albeit at the expense of using a detector with a relatively large RF bandwidth (broader than the measured absorption line). In order to efficiently down-convert the optical frequencies to the RF domain, we finally explore the possibility of implementing a dual-comb setup with a pair of FSLs. With this second technique, molecular spectroscopy is achieved with a downconversion factor >1000. In the conclusions, the prospects of acousto-optic frequency combs for spectroscopic and metrology applications, along with their limitations, are briefly discussed.

## 2. Characterization of acousto-optic FCs

The generation of acousto-optics FCs is based on a fiber frequency-shifting loop (FSL) [Fig. 1(a)]. In the first implementation described in the manuscript, the loop includes two acousto-optic frequency shifters (AOFS 1 and AOFS 2) driven by RF waveform generators. They induce frequency shifts with opposite signs (an upshift of 80 MHz and a downshift of 79 MHz). As a consequence, every roundtrip, the light frequency is shifted by $f_s = 1$ MHz. The fiber FSL also includes an Er-doped fiber amplifier (EDFA) to compensate for the loop losses, an isolator, and a tunable band-pass filter (BPF). The purpose of this filter is to control the spectral bandwidth of the comb, and to reduce the amplified spontaneous emission (ASE) generated by the EDFA. The FSL roundtrip time is close to 100 ns. The FSL is seeded by a narrow linewidth (<1 kHz) laser emitting at 1550 nm. Two 90:10 fiber couplers are employed, respectively to seed the FSL and to extract a fraction of the light circulating inside the loop, following a configuration similar to that reported in [37]. In order to visualize the lines of the generated frequency comb, we perform self-heterodyne interferometry: the light coming from the loop is mixed with the seed laser (acting as a coherent local oscillator, LO) and the beating signal is detected by a photodiode (PD). The light power in both arms of the interferometer is boosted by means of two additional EDFAs [not shown in Fig. 1(a)]. The photocurrent is digitized by an 8-bit oscilloscope and analyzed offline. The RF spectrum of the detected signal contains the beat notes between every comb line with the seed laser. A third acousto-optic modulator (AOFS 3), inducing a frequency shift of 79.5 MHz, is added to the interferometer, in order to separate the beatings of interest (i.e. the ones obtained by mixing with the seed laser) from the intra-mode beatings (i.e. the ones due to the beatings of the comb with itself).

A typical optical frequency comb produced by the FSL is shown in Fig. 1(b). This spectrum is obtained for an acquisition time of 200 μs. Since the net frequency shift is positive, high RF frequencies correspond to high optical ones. The frequency comb shown in Fig. 1(b) contains around 1800 lines within 10 dB. A zoom-in view of the comb showing individual lines at the central part of the spectrum (around 0.9 GHz) can be observed in the right plot. As expected, the frequency comb has a line spacing given by $f_s = 1$ MHz. Here, the total spectral bandwidth is limited by the BPF but, in return, the power spectrum (in dB) undergoes a roughly linear decrease of only 0.56 dB/100 lines. This slope can be adjusted to increase the signal-to-noise ratio (SNR) of the comb lines at the beginning of the spectrum, at the expense of limiting the usable bandwidth. Notice that inserting in the loop a single AOFS, which typically introduces frequency shifts of tens or a few hundreds of MHz, the measurement of the acousto-optic FC by heterodyning with the seed laser would require a significantly larger detection bandwidth. An alternative technique would involve a high-resolution optical spectral analyzer. Such a measurement, conducted in [39], demonstrates an acousto-optic FC with an optical bandwidth of 96 GHz and 1200 lines.

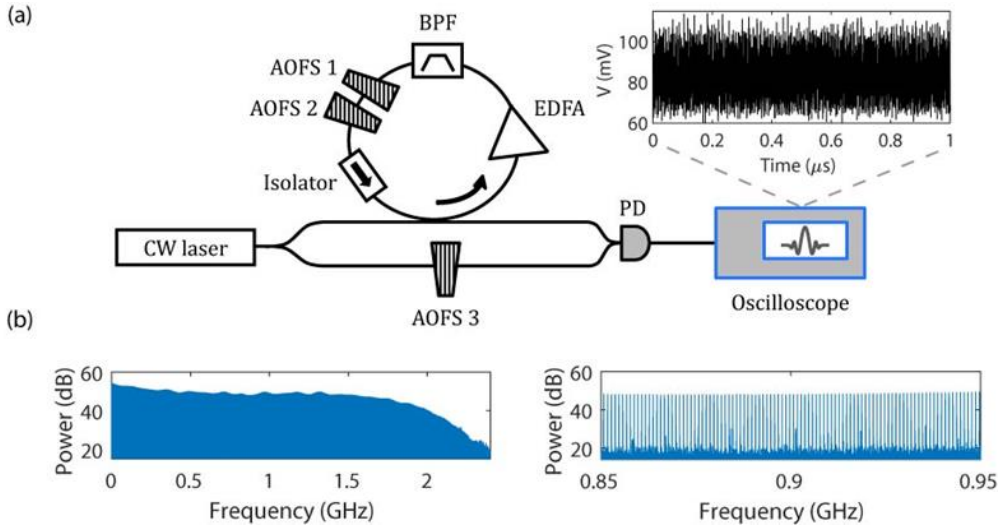

Fig. 1. (a) Sketch of the experimental setup used to generate and resolve the individual lines of an acousto-optic frequency comb by self-heterodyne interferometry. The inset shows a 1-µs fraction of a complete recorded trace. (b) RF spectrum obtained by self-heterodyning showing 1800 lines within 10 dB. The right inset is a zoom-in view of the central part of the spectrum, showing 100 lines separated from each other by 1 MHz.

## 3. Self-heterodyne spectroscopy

### 3.1 Self-heterodyne spectroscopy with sub-MHz spectral resolution

The setup displayed in Fig. 1 can generate frequency combs with an FSR easily reconfigurable, from 20 MHz (the maximum frequency difference between the two AOFSs), down to the sub-MHz frequency range. In order to demonstrate the capability of the system for high spectral resolution spectroscopy, we insert after the FSL a Fabry-Pérot cavity with a FSR of 1.6 GHz and a finesse of 250. When the line spacing is set to 5 MHz, successive transmission peaks are clearly observed in the RF frequency comb (left plot in Fig. 2, dark blue curve). This spectrum is retrieved from the Fourier transformation of a 1-ms temporal trace. Here, it is limited by the bandwidth of the oscilloscope used as digitizer. Apart from the longitudinal resonances, which are separated by the FSR, transverse modes arising from spatial mode mismatch between the laser beam and the cavity mode, are also visible in the spectrum. To determine the cavity transmission function, the comb spectrum is normalized thanks to a reference measurement performed without the resonator (left plot in Fig. 2, light blue spectrum). Due to the interferometric nature of this spectrometer, we can also retrieve the spectral phase, obtained as the difference of the phase profiles obtained with and without the cavity. The complex response of the cavity around a single resonance is shown in Fig. 2 (central plot), where the points correspond to the values retrieved at the comb lines frequencies. Notice that a comb FSR equal to 5 MHz is too large for properly resolving (in both power and phase) the transmission peak. This limitation can be solved by reducing the frequency offset between the AOFSs down to 500 kHz. This configuration provides a zoom-in view of the complex amplitude around the resonance (green area), as is shown in Fig. 2 (right plot). In this case, the transmission and phase profiles around the cavity resonance peak are clearly visible.

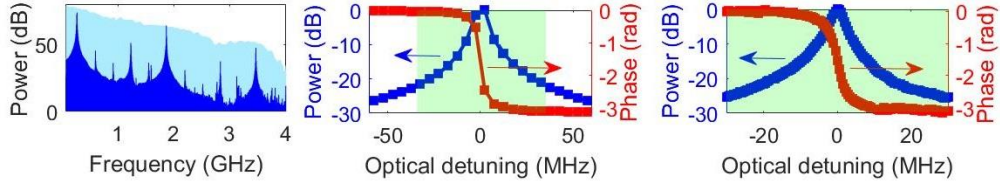

Fig. 2. Left. Dark blue: transmission spectrum of the Fabry-Pérot cavity (FSR=1.6 GHz). Light blue: reference spectrum (no cavity inserted). The line spacing is 5 MHz. Center: Complex amplitude of a single transmission peak. The phase curve shows the expected $\pi$-jump. Right: Zoom-in view of the transmission peak (green area in the central plot) when the line spacing is reduced to 500 kHz.

*3.2 Self-heterodyne molecular spectroscopy*

In a second set of experiments, we apply self-heterodyne interferometry to conduct fast molecular spectroscopy. We use the experimental setup shown in Fig. 3(a), which includes a FSL with a single frequency shifter (AOFS1) to generate a comb with a line spacing of 80 MHz. With this setup, we report the measurement of an absorption line of hydrogen cyanide ($H^{13}C^{14}N$), a common wavelength reference in the C-telecommunications band [4, 18, 19]. To this end, we place a fiber-coupled HCN cell (pressure: 25 Torr, length: $L = 5.5$ cm) after the FSL. Here, the CW laser is replaced with a diode laser tunable around 1550 nm and with a linewidth < 500 kHz. This laser allows us to measure the complex amplitude of the 11$^{th}$ line in the HCN P-branch (line P11). In order to obtain a reference measurement (without cell), a 3dB coupler splits the signal arm into two arms, one containing the fiber gas cell, and the other one a frequency shifter inducing a negative shift (AOFS2, driven at 87 MHz). By means of two 3dB couplers, the light coming from both arms is mixed with the LO [17]. As before, the light frequency in the LO arm is shifted by a third acousto-optic modulator (AOFS3, driven at 78 MHz), to separate the intra-mode beat notes from the spectrum of interest. The power of both interferometer arms are boosted by two additional EDFAs [not shown in the Fig. 3(a)]. Apart from the intra-mode beat notes, the RF spectrum of the photodiode signal is composed of two interleaved combs: one corresponds to the absorption profile, and the other one to the reference measurement. Since self-heterodyne interferometry provides a direct mapping of the optical spectrum into the RF domain (i.e., the spectral scaling factor, or compression factor, is 1), the measurement of the whole molecular absorption line requires the use of a broadband oscilloscope (bandwidth >8 GHz).

In order to measure the molecular transition, a set of 25 temporal traces is recorded. Each one has a duration of 20 $\mu s$. The complex amplitude (modulus and phase) retrieved after a baseline correction, is shown in Fig. 3(b), where each point is the result of averaging the values obtained from all traces. The transmission curve is satisfactorily fitted by a Voigt profile, and the theoretical phase is subsequently derived from the fitted amplitude using the Kramers-Kronig relations. The root mean square (rms) of the residuals between the experimental transmission curve and the Voigt profile is $\sigma = 0.8$ %, at the noise level. In the low absorption regime, this value sets a limit on the absorbance $\alpha L$ that can be measured for an acquisition time of 500 µs. From $\sigma$, one can estimate the noise-equivalent absorption sensitivity (NEAS), defined as $NEAS = \sigma/(L \sqrt{BW})$, where $BW$ is the detection bandwidth [41]. For our self-heterodyne spectrometer, $BW = 8$ GHz and then $NEAS = 1.6\times10^{-8}$ cm$^{-1}$ Hz$^{-1/2}$ ( 500 µs time-averaging). Assuming that $\sigma$ scales as the inverse of the root-square of the acquisition time, the value of the NEAS for 1-s time averaging would be comparable to the one reported in [19].

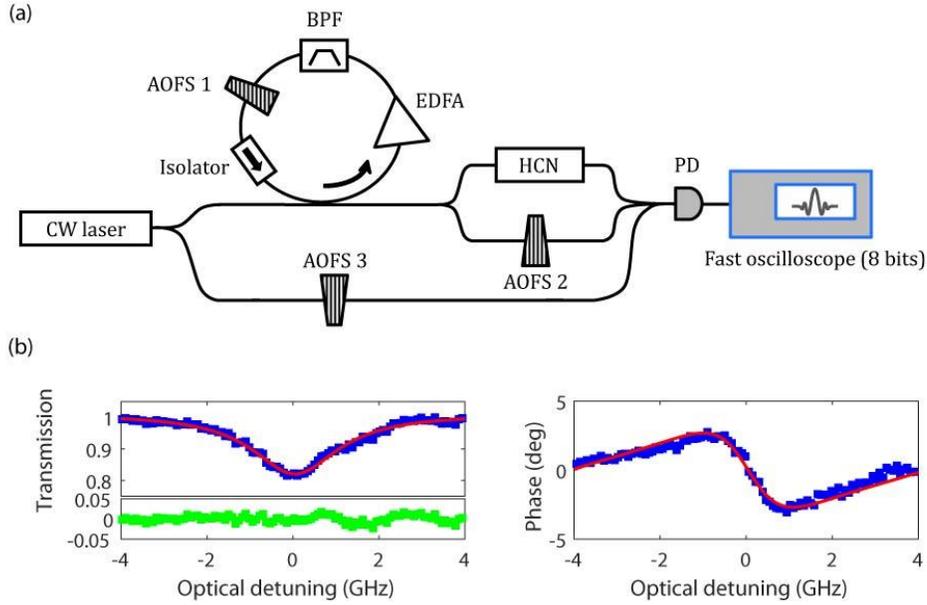

Fig. 3. (a) Sketch of the experimental setup used to perform self-heterodyne molecular spectroscopy of HCN. (b) Transmission and phase corresponding to the 11[th] line of the HCN P-branch. In the transmission curve, the red line is the Voigt profile that fits the experimental data (in blue) and the green squares plotted below are the corresponding residuals. In the phase plot, the red curve corresponds to the theoretical function derived from the Voigt profile.

## 4. Dual-comb spectroscopy

### 4.1 Experimental setup

Practically, self-heterodyne interferometry is limited by the fact that the bandwidth of the detection chain must match the optical bandwidth of the FC. This requirement can be circumvented using dual-comb interferometry, which enables to perform multi-heterodyne spectroscopy with a much narrower detection bandwidth [4, 42]. Instead of making a probe comb (with a line spacing $f_p$) interfere with a single frequency laser, a second comb with a slightly different line spacing ($f_{LO} = f_p + \delta f$) is used as a LO [see Fig. 4(a)]. The two combs are mixed in a photodetector and the application of a low-pass filter on the photocurrent provides beat notes between every line of the probe comb and the closest line of the LO. The resulting spectrum is an RF comb, with a line spacing of $\delta f$, which is orders of magnitude lower than $f_p$, ($\delta f \ll f_p$). This frequency downconversion is characterized by a compression factor defined as: $C = f_p/\delta f \gg 1$. The frequency response of a sample, encoded on the lines of the probe comb, can be thus measured using a signal with a RF bandwidth: $BW = BW_{opt}/C$, where $BW_{opt}$ is the optical bandwidth of the probe comb (i.e. the number of comb lines multiplied by the line spacing). The minimum time required to perform a complete measurement (i.e., the duration of an interferogram) is $\tau_0 = 1/\delta f$. In practice, the acquisition time $\tau_{acq}$ is usually much longer than $\tau_0$, in order to increase the signal-to-noise ratio [42].

A critical point for resolving each line of the RF spectrum is to ensure a high degree of mutual coherence between the combs. Similarly to electro-optic dual-comb spectrometers, in acousto-optic frequency combs, both FSLs are seeded by the same CW laser, a configuration much simpler than the one involving two stabilized frequency combs. However, when the acquisition time becomes too long (in order to average thousands of interferograms), small perturbations of the relative optical path-length between the two FSLs (due, for instance, to mechanical vibrations or thermal drifts) limit drastically the bandwidth of the RF spectrum. This limitation is discussed in the conclusion paragraph. To overcome this restriction and provide a mutual stabilization of the two FSLs, we implemented a low-bandwidth control loop [in green in Fig. 4(a)]. An error signal is generated by mixing one line of the RF spectrum (25$^{th}$ line), obtained from the beating of the probe comb with the LO, with a sine wave provided by an external function generator at the same frequency. This generator is locked to the clock that drives the acousto-optic shifters in the FSLs (system clock). The error signal is amplified and low-pass filtered (cutoff frequency at 300 Hz), before being sent to the PID (proportional-integral-derivative) controller. Then, the signal generated by the PID controller is applied to an electro-optic phase modulator (PM) inserted in the probe FSL. The PM acts like a tunable delay line, which enables to cancel in real-time the relative length fluctuations of the two FSLs. To demonstrate the performance of the control loop, we use a simplified version of the setup shown in Fig. 4(a), where the output of the probe comb directly interferes with the output of the LO comb. The offset between the two combs is 50 kHz ($C$=1600) and both FSLs are fed by the laser used in Sec. 2 (1550 nm). The lower inset in Fig. 4(a) shows the Fourier transform of a 50-ms trace (containing 2500 interferograms) with and without the PID regulation (respectively blue and red spectra). A zoom-in view around the line 180 is also included to evidence the effect of the PID regulator. In practice, the stabilization scheme enables to use about 240 mutually coherent lines, with relative intensity variations lower than 10 dB, within 50 ms.

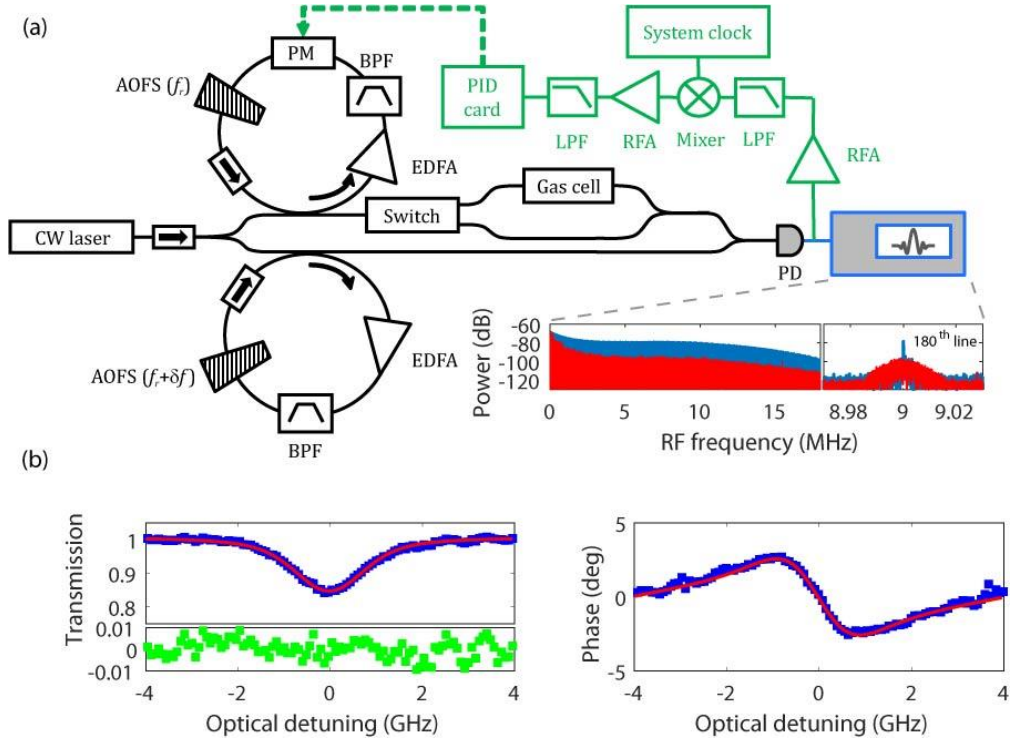

Fig. 4. (a) Sketch of the optical setup used to perform dual-comb spectroscopy. RFA: RF amplifier; LPF: low-pass filter. Lower inset: RF spectrum obtained by directly mixing the output of the two FSLs with and without the stabilization loop (blue and red spectra, respectively). A zoom-in view of a single line is also included. The PID regulator allows obtaining around 240 lines within 10 dB. (b) Transmission and phase corresponding to the 11$^{th}$ line of the HCN P-branch. In the transmission curve, the red line is the Voigt profile that fits the experimental data (in blue) and the green squares plotted below are the corresponding

residuals. In the phase plot, the red curve corresponds to the theoretical function derived from the Voigt profile.

*4.2 Molecular spectroscopy results*

To prove the capability of acousto-optic FCs for dual-comb spectroscopy, we provide an experimental measurement of the P11 HCN absorption line. The experimental setup is shown in Fig. 4(a). To this end, we insert the HCN cell in one arm of the interferometer. The light source that feeds the two comb generators is the tunable laser employed in Sec. 3.2. The offset between the two combs is: $\delta f = 60$ kHz and the compression factor is around 1300. A fiber pigtailed optical switch is inserted in the probe arm, to take a reference measurement, in such a way that every trace registered by the oscilloscope contains at least a pair of consecutive temporal signals, respectively recorded with and without the gas cell. In our experiments, the total duration of each recorded trace is 200 ms, from which two 50 ms long signals are extracted (one corresponding to the probe and the other one to the reference). Fig. 4(b) shows the retrieved complex response corresponding to the P11 line, after applying a baseline correction algorithm, which removes the systematic variations between the signal and the reference. The blue points are the result of averaging 20 independent traces. The results are very similar to those obtained by means of self-heterodyne spectroscopy. A theoretical transmission Voigt profile and the theoretical phase derived by the Kramers-Kronig equations are also included in Fig. 4(b). The error in the transmission curve is $\sigma = 0.4$ % at the noise level, leading to a NEAS at 1s time-averaging of $3\times10^{-7}$ cm$^{-1}$ Hz$^{-1/2}$, a value somewhat lower than the one reported in [33].

## 5. Conclusions and discussion

We have demonstrated a new and versatile source of optical FCs, based on repetitive frequency shifts of a single-frequency laser in a FSL. This technique provides a flat optical comb composed of >1500 lines (within 10 dB). The line spacing is easily reconfigurable and ranges from tens of MHz down to the sub-MHz range. In a first implementation, the FSL included a pair of acousto-optic modulators inducing frequency shifts with opposite signs. Using this configuration, we have been able to record, by self-heterodyne interferometry, the transmission function of a Fabry-Pérot cavity with a spectral resolution of hundreds of kHz. When only a single acousto-optic modulator is inserted in the FSL, the comb bandwidth can be extended well beyond the GHz range. Using again a self-heterodyne technique and a fast detection chain, we have measured an absorption line of HCN at 1550 nm, covering an optical bandwidth of 8 GHz. This measurement has been carried out on the sub-millisecond time scale, a speed that makes unnecessary the use of control loops to stabilize the fiber interferometer. In order to eliminate the need for a large-bandwidth detection chain, we have implemented a multi-heterodyne interferometer with a pair of frequency combs generated in two different FSLs. This dual-comb scheme has provided a down-conversion factor larger than 1000 by setting a frequency offset between the combs of tens of kHz. To ensure the mutual coherence of the two combs on the measurement time scale, a PID control loop has been employed to compensate for the fluctuations in the relative optical path length between the two combs. This setup has been used successfully for the measurement of a HCN absorption line.

In conclusion, we have proven experimentally the capability of acousto-optic FCs to perform multi-heterodyne spectroscopy in different interferometric configurations. This comb modality presents many advantages. Similarly to combs generated by electro-optic modulation, the absolute comb frequencies are set by the frequency of the CW laser and the comb spacing is not linked to the length of a cavity. Moreover, a pair of mutually coherent FCs can be generated from the same CW laser. Finally, unlike previous approaches that involve non-linear optics phenomena or fast electronics to broaden the spectra, our approach permits the generation of combs with > 1500 coherent lines, and this without adding to the system hardware and/or software complexity. The high spectral resolution of acousto-optic FCs (sub-MHz), combined with the large number of spectral lines, makes them suitable for optical fiber sensing [29] or real-time atomic spectroscopy [30].

Contrary to electro-optic FCs, where all spectral lines are produced simultaneously in the electro-optic modulator, the different frequencies of the acousto-optic FCs are created by successive frequency shifts in a recirculating loop. Therefore, in order to perform coherent self-heterodyne interferometry, the coherence time of the seed laser must exceed the loop round-trip time multiplied by the total number of lines. Considering a FSL round-trip time of 100 ns and a number of lines of $N = 1000$, this condition requires to seed the FSL with a CW laser having a coherence time longer than 100 µs (i.e. a linewidth below 10 kHz). This condition is much less stringent in dual-comb interferometry: in that technique what matters is the mutual coherence between the consecutive pairs of lines generated by the two combs, which is easily satisfied if the two FSLs have comparable lengths. Apart from the requirements on the laser linewidth, in standard experimental conditions FSLs undergo mechanical disturbances arising from vibrations and thermal drifts. In general, the frequencies of mechanical noise are below 10 kHz, which means that, with the above parameters, the loop can be considered to be fixed on the time scale of $N$ roundtrips. However, if the acquisition time exceeds by far this time scale, random fluctuations of the FSL length result in fluctuations of the comb spectral phase, and thus in a broadening of the comb lines. . In this case, locking the length of the FSLs by means of a control loop is required, as in the dual-comb technique reported above. Notice that stabilization schemes more sophisticated than the one included in our setup can potentially increase the number of useable lines, as well as allow us to accumulate interferograms for hours [27]. Finally, the ultimate performance of acousto-optics FCs is linked to the maximum number of lines that can be created in a FSL. Here, we have reported the generation of more than 1500 mutually coherent lines. It turns out that this number depends critically on parameters such as the power of the seed laser, the gain and the noise factor of the EDFA, and the passive losses of the FSL [40]. An extensive investigation of the influence of these parameters, and the effects of gain saturation on the number of lines, will be published shortly.

## Acknowledgments

We acknowledge support from the Agence Nationale de la Recherche (grant #ANR-14-CE32-0022). We also thank Tektronix France for the loan of a 25 GHz-bandwidth real-time digital oscilloscope.

## References and links